\documentclass[final,5p,times,twocolumn]{elsarticle}
\pdfoutput=1
\usepackage{amsmath}

\usepackage{amssymb}
\usepackage{bm}% bold math
\renewcommand{\vec}[1]{\boldsymbol{#1}}
\newcommand{\be}{\begin{equation}}
\newcommand{\ee}{\end{equation}}
\newcommand{\bea}{\begin{eqnarray}}
\newcommand{\eea}{\end{eqnarray}}
\newcommand{\Tr}{\,\hbox{\rm Tr}}

\newcommand{\msbar}{{\overline{\mbox{\scriptsize MS}}}}

\journal{CERN-TH-2016-211}

\begin{document}

\begin{frontmatter}

%% Title, authors and addresses

%% use the tnoteref command within \title for footnotes;
%% use the tnotetext command for theassociated footnote;
%% use the fnref command within \author or \address for footnotes;
%% use the fntext command for theassociated footnote;
%% use the corref command within \author for corresponding author footnotes;
%% use the cortext command for theassociated footnote;
%% use the ead command for the email address,
%% and the form \ead[url] for the home page:
%% \title{Title\tnoteref{label1}}
%% \tnotetext[label1]{}
%% \author{Name\corref{cor1}\fnref{label2}}
%% \ead{email address}
%% \ead[url]{home page}
%% \fntext[label2]{}
%% \cortext[cor1]{}
%% \address{Address\fnref{label3}}
%% \fntext[label3]{}
  
\title{Equation of state of the SU($3$) Yang--Mills theory:\\
a precise determination from a moving frame}

\author[a,b,c]{Leonardo Giusti} 
\author[c]{Michele Pepe}

\address[a]{Theoretical Physics Department, CERN, Geneva, Switzerland}

\address[b]{Dipartimento di Fisica, Universit\`a di Milano-Bicocca\\ Piazza della Scienza 3, 
  I-20126 Milano, Italy}

\address[c]{INFN, Sezione di Milano-Bicocca \\ Piazza della Scienza 3, 
I-20126 Milano, Italy}

\begin{abstract}
The equation of state of the SU($3$) Yang--Mills theory is determined
in the deconfined phase with a precision of about 0.5\%.
The calculation is carried out by numerical simulations of lattice gauge
theory with shifted boundary conditions in the time direction. At each given
temperature, up to $230\, T_c$ with $T_c$ being the critical temperature,
the entropy density is computed at several lattice spacings so to be able to
extrapolate the results to the continuum limit with confidence. Taken at face value,
above a few $T_c$ the results exhibit a striking linear behaviour in $\ln(T/T_c)^{-1}$
over almost 2 orders of magnitude. Within errors, data point straight to the
Stefan-Boltzmann value but with a slope grossly different from the leading-order
perturbative prediction. The pressure is determined by integrating the entropy in the
temperature, while the energy density is extracted from $T s=(\epsilon + p )$. The
continuum values of the potentials are well represented by Pad\'e interpolating formulas,
which also connect them well to the Stefan-Boltzmann values in the infinite
temperature limit. The pressure, the energy and the entropy
densities are compared with results in the literature. The discrepancy among
previous computations near $T_c$ is analyzed and resolved thanks to the high precision
achieved. 
\end{abstract}

%\begin{keyword}

  %% keywords here, in the form: keyword \sep keyword

%% PACS codes here, in the form: \PACS code \sep code

%% MSC codes here, in the form: \MSC code \sep code
%% or \MSC[2008] code \sep code (2000 is the default)

%\end{keyword}

\end{frontmatter}

%% \linenumbers

\section{Introduction}\label{intro}
The equation of state (EoS) of strongly-interacting 
matter is of absolute interest in particle and nuclear
physics, and in cosmology. Apart the obvious theoretical interest
in such a basic property, the collective
behaviour of strongly-interacting particles has
determined the evolution of the Universe in its early stages.
Those extreme conditions are now being reproduced
and investigated at heavy-ion colliders, where
the EoS is a crucial input in the analysis of data.\\
\indent Lattice gauge theory is the theoretical framework where the EoS
is determined from first principles.
A first measurement in the SU($3$) Yang-Mills theory was
performed in Ref.~\cite{Boyd:1996bx}. The pressure $p$, the entropy density $s$
and the energy
density $\epsilon$ were computed with a numerical accuracy of
about 1-2\%  up to temperatures of $\sim\!\! 5 T_c$. The strategy that was used, 
the ``integral'' method\footnote{For a variant see Ref.~\cite{Umeda:2008bd}.},
has become the most popular technique in numerical
investigations of the EoS. It is based on the direct
measurement of the trace anomaly
by Monte Carlo simulations. The pressure is then obtained
by integrating in the temperature, while the entropy and the energy densities
are calculated using the thermodynamic relation $T s=(\epsilon + p )$.
In Ref.~\cite{Borsanyi:2012ve} a refined version of the integral method was used
to determine the EoS in a much broader range of temperatures, from $0$ up to
$\sim\! 1000\, T_c$, with a target accuracy at the permille level. In the region
near $T_c$, the results of the two computations show significant discrepancies.\\ 
\indent Severe limitations hinder the integral method.
The need for the (zero temperature) subtraction of the ultraviolet power divergence,
and the complicated procedure for determining the lines of constant
physics make the computation very demanding numerically and 
technically involved, see Refs.~\cite{Philipsen:2012nu,Ding:2015ona}
for recent reviews. Despite the impressive progress over the last few
years~\cite{Borsanyi:2013bia,Bazavov:2014pvz}, uncertainties in the EoS of
Quantum Chromodynamics (QCD) are still
rather large, and temperatures higher than a few hundreds MeV are still
unreachable with staggered fermions. The computation remains prohibitive
with Wilson fermions.\\
\indent These obstacles are not rooted in the physics of the EoS, 
but in the method adopted for its computation.  Recently
there has been an intense activity to design new numerical
strategies for simulating thermal field theories on the lattice and in particular
to compute the
EoS~\cite{Giusti:2010bb,Giusti:2011kt,Giusti:2012yj,Asakawa:2013laa,Giusti:2014ila,Umeda:2014ula,Taniguchi:2016ofw,Kitazawa:2016dsl}.\\
\indent Among the proposed new methods, we have been focusing on the formulation of a
thermal theory in a moving reference frame~\cite{Giusti:2010bb,Giusti:2011kt,Giusti:2012yj}.
In this approach the entropy density is the primary observable computed by Monte Carlo
simulations. It is extracted from the expectation
value of the space-time component of the energy-momentum tensor $T_{0k}$
in presence of shifted boundary conditions~\cite{Giusti:2012yj,Giusti:2014ila}.
This quantity has no ultraviolet power divergences, and the finite multiplicative
renormalization constant can be computed by imposing suitable Ward
identities~\cite{Giusti:2012yj,Giusti:2015daa}. The pressure is then computed by integrating
the entropy in the temperature, while the energy density is determined by integrating the
temperature in the entropy or equivalently by using the relation
$T s=(\epsilon + p )$.\\
\indent Following this strategy, in the last few years we have carried out
extensive numerical computations in the SU($3$) Yang--Mills theory also
to prepare the ground for QCD. It is the aim of this letter to present the
final results of this study.

\section{Preliminaries}
We regularize the SU($3$) Yang--Mills theory on a finite four-dimensional
lattice of spatial volume $V=L^3$, temporal direction $L_0$, and spacing $a$.
The gauge field satisfies periodic boundary conditions in the 
spatial directions and shifted boundary conditions 
in the temporal direction 
\be
U_\mu(L_0,\vec x) =U_\mu(0,\vec x - L_0\vec\xi)\; ,
\ee
where $U_\mu(x_0,\vec x)\in$ SU($3$) are the link variables, and
the spatial vector $\vec \xi$ characterizes the moving frame
in the Euclidean space-time~\cite{Giusti:2010bb,Giusti:2011kt,Giusti:2012yj}. 
The action is discretized through the standard Wilson plaquette 
\be\label{eq:latS}
S[U] = \frac{3}{g_0^2}\, \sum_{x} \sum_{\mu,\nu} 
\left[1 - \frac{1}{3}{\rm Re}\Tr\Big\{U_{\mu\nu}(x)\Big\}\right]\; ,
\ee
where the trace is over the color index, and $g_0$ is the bare coupling
constant. The plaquette, defined as a function 
of the gauge links, is
\be\label{eq:placst}
U_{\mu\nu}(x) = U_\mu(x)\, U_\nu(x+ a\hat \mu)\, U^\dagger_\mu(x + a\hat \nu)\,
             U^\dagger_\nu(x)\; ,  
\ee
where $\mu,\,\nu=0,\dots,3$, $\hat \mu$ is the unit vector along the 
direction $\mu$, and $x$ is the space-time coordinate. We are interested
in the off-diagonal components of the energy-momentum tensor
\be\label{eq:tmunuT}
T_{0k} = \frac{1}{g^2_0}\, F^a_{0\alpha}F^a_{k\alpha} \; ,
\ee
where the gluon field strength tensor is defined as~\cite{Caracciolo:1989pt} 
\be
F^a_{\mu\nu}(x) = - \frac{i}{4 a^2} 
\Tr\Big\{\Big[Q_{\mu\nu}(x) - Q_{\nu\mu}(x)\Big]T^a\Big\}\; , 
\ee
and 
$Q_{\mu\nu}(x) = U_{\mu\nu}(x) + U_{\nu-\mu}(x) + U_{-\mu-\nu}(x) + U_{-\nu\mu}(x)$,
the minus sign standing for negative orientation. The
$T^a$ are the Hermitian generators, see Ref.~\cite{Giusti:2015daa}
for additional details. 

\subsection{The renormalization constant $Z_T$}
The field $T_{0k}$ is multiplicatively renormalized
by the finite renormalization constant $Z_T(g^2_0)$
which we have computed non-perturbatively in
Ref.~\cite{Giusti:2015daa}. To have additional control
on its discretization effects,
we have also calculated the very same quantity at one loop
in perturbation theory
at finite lattice spacing~\cite{GiustiPepe2016PT}. By re-analyzing the data
in Ref.~\cite{Giusti:2015daa} with the help of the one-loop improved definition,
the final results for $Z_T(g^2_0)$
are well represented by the expression
\bea
Z_{_T}(g_0^2) & = & \frac{1 - 0.4367\, g_0^2}{1 - 0.7074\, g_0^2} - 0.0971\, g_0^4\nonumber\\[0.25cm]
& + & 0.0886\, g_0^6 - 0.2909\, g_0^8
\eea
in the full range $0\leq\! g^2_0\!\leq 1$. The error to be attached to 
this function, computed as in Ref.~\cite{Giusti:2015daa}, is $0.4\%$ up 
to $g_0^2\leq 0.85$ while it grows linearly from $0.4\%$ to $0.7\%$ 
in the range $0.85\leq g_0^2\leq 1$. This new determination of $Z_{_T}(g_0^2)$
is within one standard deviation from the one in Ref.~\cite{Giusti:2015daa},
a fact which confirms that the systematics due to discretization
effects is well within the quoted error.
\begin{figure}[t!]
\includegraphics[width=.50\textwidth]{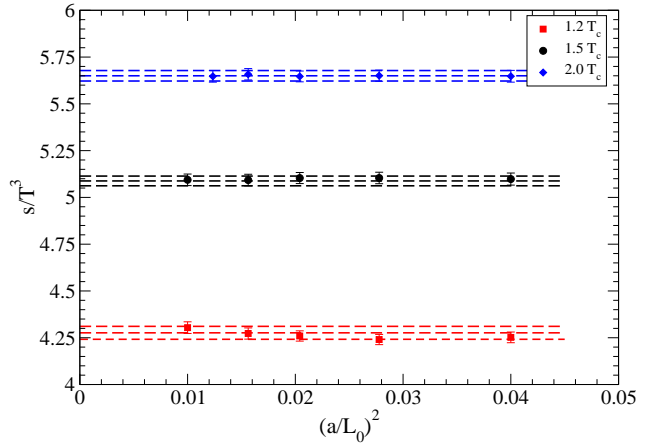}
\caption{
Entropy density $s/T^3$ versus $(a/L_0)^2$ for $T=1.2, 1.5, 2.0$~$T_c$.
\label{plotCL}}
\end{figure}
\begin{table*}[t!]
\begin{center}
\setlength{\tabcolsep}{.30pc}
%\begin{tabular}{@{\extracolsep{0.4cm}}cccc|cccc|cccc}
\begin{tabular}{|lccc|lccc|lccc|}
\hline%\\[-0.175cm]
& & & & & & & & & & & \\[-0.25cm]  
$T/T_c$  &$s/T^3$&$p/T^4$&$\epsilon/T^4$&
$T/T_c$  &$s/T^3$&$p/T^4$&$\epsilon/T^4$&
$T/T_c$  &$s/T^3$&$p/T^4$&$\epsilon/T^4$\\[0.125cm]
\hline%\\[-0.175cm]
$0.660$ & $0.00(4)$   &$0.00004(15)$& $0.0010(28)$ & $1.807$ & $5.481(27)$ & $1.125(7)$ & $4.355(20)$ & $7.228$ & $6.462(26)$ & $1.592(6)$ & $4.871(19)$ \\[0.125cm]
$0.904$ & $0.108(23)$ & $0.006(4)$  & $0.104(27)$ & $1.900$ & $5.570(27)$ & $1.172(7)$ & $4.398(20)$ & $10.22$ & $6.540(26)$ & $1.619(6)$ & $4.921(20)$ \\[0.125cm]
$0.980$ & $0.31(11) $ & $0.018(5)$  & $0.28(11)$ & $2.000$ & $5.650(28)$ & $1.215(7)$ & $4.435(21)$ & $14.46$ & $6.595(26)$ & $1.638(7)$ & $4.957(20)$ \\[0.125cm]
$1.061$ & $3.22(15) $ & $0.149(10)$  & $3.07(14)$ & $2.300$ & $5.80(4)$   & $1.309(8)$ & $4.49(3)$ & $20.44$ & $6.630(27)$ & $1.650(7)$ & $4.980(20)$ \\[0.125cm]
$1.100$ & $3.60(3)$   & $0.243(11)$ & $3.361(24)$ & $2.556$ & $5.903(24)$ & $1.362(8)$ & $4.541(17)$ & $28.91$ & $6.670(27)$ & $1.660(7)$ & $5.010(20)$ \\[0.125cm]
$1.150$ & $4.00(4)$   & $0.359(10)$ & $3.64(3)$ & $2.711$ & $5.969(25)$ & $1.388(7)$ & $4.581(17)$ & $40.89$ & $6.715(27)$ & $1.671(7)$ & $5.045(20)$ \\[0.125cm]
$1.200$ & $4.28(3)$   & $0.465(10)$ & $3.811(25)$ & $2.891$ & $6.017(25)$ & $1.413(7)$ & $4.604(18)$ & $57.82$ & $6.738(27)$ & $1.680(7)$ & $5.058(20)$ \\[0.125cm]
$1.278$ & $4.554(26)$ & $0.607(9)$  & $3.947(18)$ & $3.072$ & $6.065(25)$ & $1.434(7)$ & $4.631(18)$ & $81.78$ & $6.745(27)$ & $1.684(7)$ & $5.061(20)$ \\[0.125cm]
$1.400$ & $4.907(27)$ & $0.785(8)$  & $4.122(19)$ & $3.253$ & $6.108(24)$ & $1.452(7)$ & $4.656(18)$ & $115.6$ & $6.764(27)$ & $1.688(7)$ & $5.076(20)$ \\[0.125cm]
$1.500$ & $5.098(26)$ & $0.898(8)$  & $4.201(19)$ & $3.433$ & $6.141(25)$ & $1.467(7)$ & $4.674(18)$ & $163.6$ & $6.778(27)$ & $1.692(7)$ & $5.086(20)$ \\[0.125cm]
$1.600$ & $5.252(27)$ & $0.988(8)$  & $4.264(20)$ & $3.614$ & $6.178(25)$ & $1.481(7)$ & $4.697(18)$ & $231.3$ & $6.788(27)$ & $1.695(7)$ & $5.093(20)$ \\[0.125cm]
$1.700$ & $5.368(26)$ & $1.061(7)$  & $4.307(19)$ & $5.111$ & $6.350(25)$ & $1.549(6)$ & $4.801(19)$ &         &             &             &             \\[0.125cm]
\hline
\end{tabular}
\caption{
  \label{tab:continuum} Continuum limit values of the normalized entropy density,
  pressure and energy density.}
\end{center}
\end{table*}

\section{The entropy density from the moving frame}
When $\vec\xi$ is non zero, the entropy density at 
the temperature $T=1/(L_0\sqrt{1+{\bf \vec \xi}^2})$
can be written as~\cite{Giusti:2012yj}
\begin{equation}\label{eq:entropy}
  \frac{s(T)}{T^3}= \frac{L_0^4 (1+{\bf \vec \xi}^2)^3}{\xi_k} \, \langle T_{0k}
  \rangle_{\vec \xi} \; Z_T\; . 
\end{equation}
Once $Z_T(g_0^2)$ has been determined, Eq.~(\ref{eq:entropy}) provides a simple
way for measuring the entropy density. The expectation value
$\langle T_{0k} \rangle_{\vec \xi}$ can be measured by a single Monte
Carlo simulation at each temperature and lattice spacing. No power divergences
have to be subtracted, and the continuum limit is simply attained by increasing
$L_0/a$ and by tuning the bare coupling constant $g_0$ so that the temperature
stays unchanged in physical units.
Since the observable is ultralocal, numerical
simulations of lattices with large spatial sizes are not more expensive. The additional
cost of updating a large lattice is compensated by the reduced statistical error
due to volume averaging.\\
\indent In our computation we opted for ${\vec\xi} = (1,0,0)$ for most of the
temperatures, since very small lattice artifacts have been previously observed
for the tree-level improved definition of $\langle T_{0k} \rangle_{\vec \xi}$ at
this value of the shift~\cite{Giusti:2012yj,Giusti:2014ila,Giusti:2015daa}. We
have used\footnote{For $T/T_c=0.660$ we have computed the entropy density from the
expectation value of $\langle T_{00}-T_{kk}\rangle $ without shifted boundary
conditions, and used the renormalization constant in Ref.~\cite{Giusti:2015daa}.
For $T/T_c=0.904$, $s/T^3$ has been determined by using the results at $T/T_c=1.278$
combined with the step-scaling function in Ref.~\cite{Giusti:2014ila}.}
${\vec \xi} = (1,1,0)$ for $T/T_c=0.980$, and 
${\vec \xi} = (1,1,1)$ for $T/T_c=0.904$, $1.00$, $1.061$ and $2.30$.
In order to perform the continuum limit extrapolation of
$s(T)/T^3$ with confidence, at each given temperature we have carried out numerical
simulations with $L_0/a=5$, $6$, $7$, $8$ and, sometimes, also $3$, $4$, $9$ and $10$.
The inverse coupling constant $6/g_0^2$ has been fixed as in Ref.~\cite{Giusti:2014ila}:
from the Sommer scale $r_0/a$~\cite{Necco:2001xg} up to temperatures of $2\, T_c$,
while for higher temperatures from a quadratic interpolation in $\ln{(L/a)}$ of the data listed
in Tables A.1 and A.4 of Ref.~\cite{Capitani:1998mq} corresponding to fixed values of the
Schr\"odinger functional coupling constant $\bar g^2(L)$. The critical temperature in units
of the Sommer scale is $r_0 T_c=0.750(4)$~\cite{Boyd:1996bx,Lucini:2003zr}. 
For temperatures above $2 T_c$, the accuracy on the temperature is about
$2\%$~\cite{Capitani:1998mq} but its effect is negligible due to the very weak dependence
of the thermodynamic quantities on the temperature. Error bars of the Monte Carlo data
have been estimated using both the jackknife method and the binnining technique finding
consistent results.
\begin{figure}[t!]
\includegraphics[width=0.525\textwidth]{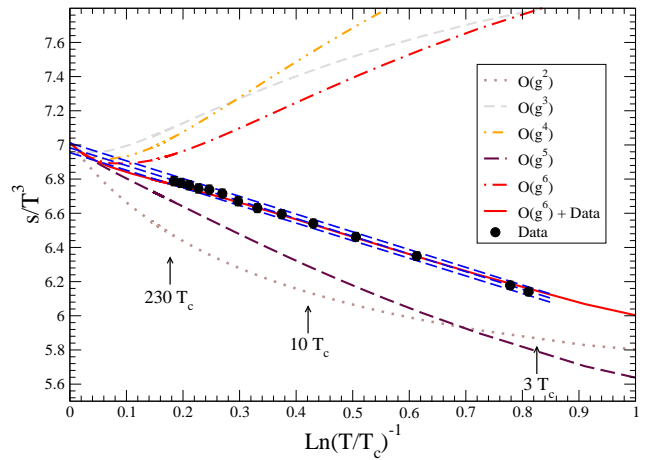}
\caption{Entropy density $s/T^3$ versus $\ln(T/T_c)^{-1}$. 
Circles are data from Table~\ref{tab:continuum} for $T\geq 3.433 T_c$, while
the dashed (blue) curves represent the linear interpolation formula in
Eq.~(\ref{eq:pades}) and its error. The other curves show the perturbative
expression in Ref.~\cite{Kajantie:2002wa}, each one including up to the order indicated in the
legend. The continuous (red) curve is the  ${\cal O}(g^6)$ perturbative prediction,
but with a non-null unknown term fixed so to be able to reproduce the data in the
temperature range considered.\label{fig:sPT}}
\end{figure}
In order to keep finite volume effects
below the statistical errors, the lattice size in the spatial directions has been chosen
to be $L/a=128$ for $L_0/a$ up to $6$ and $L/a=256$ for larger values. Thus, $L T$
ranges from $10$ to $26$, values where finite size effects are negligible within our statistical
errors~\cite{Giusti:2012yj,Giusti:2014ila}.\\
\begin{figure*}[th]
  \vspace{1.5cm}
\begin{center}
\begin{minipage}{0.48\textwidth}
\includegraphics[width=9.5 cm,angle=0]{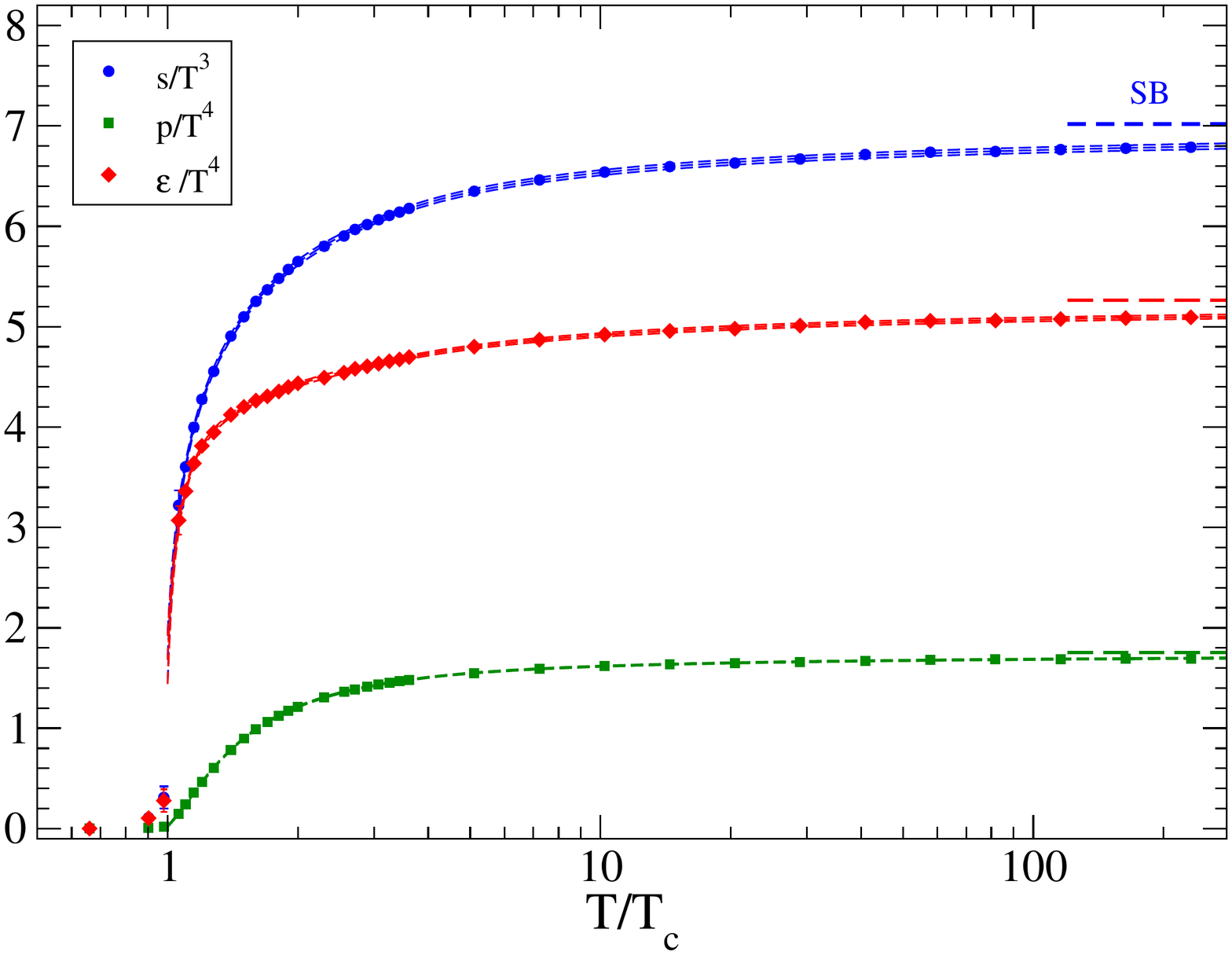}
\end{minipage}
\hspace{5mm}
\begin{minipage}{0.48\textwidth}
\includegraphics[width=9.5 cm,angle=0]{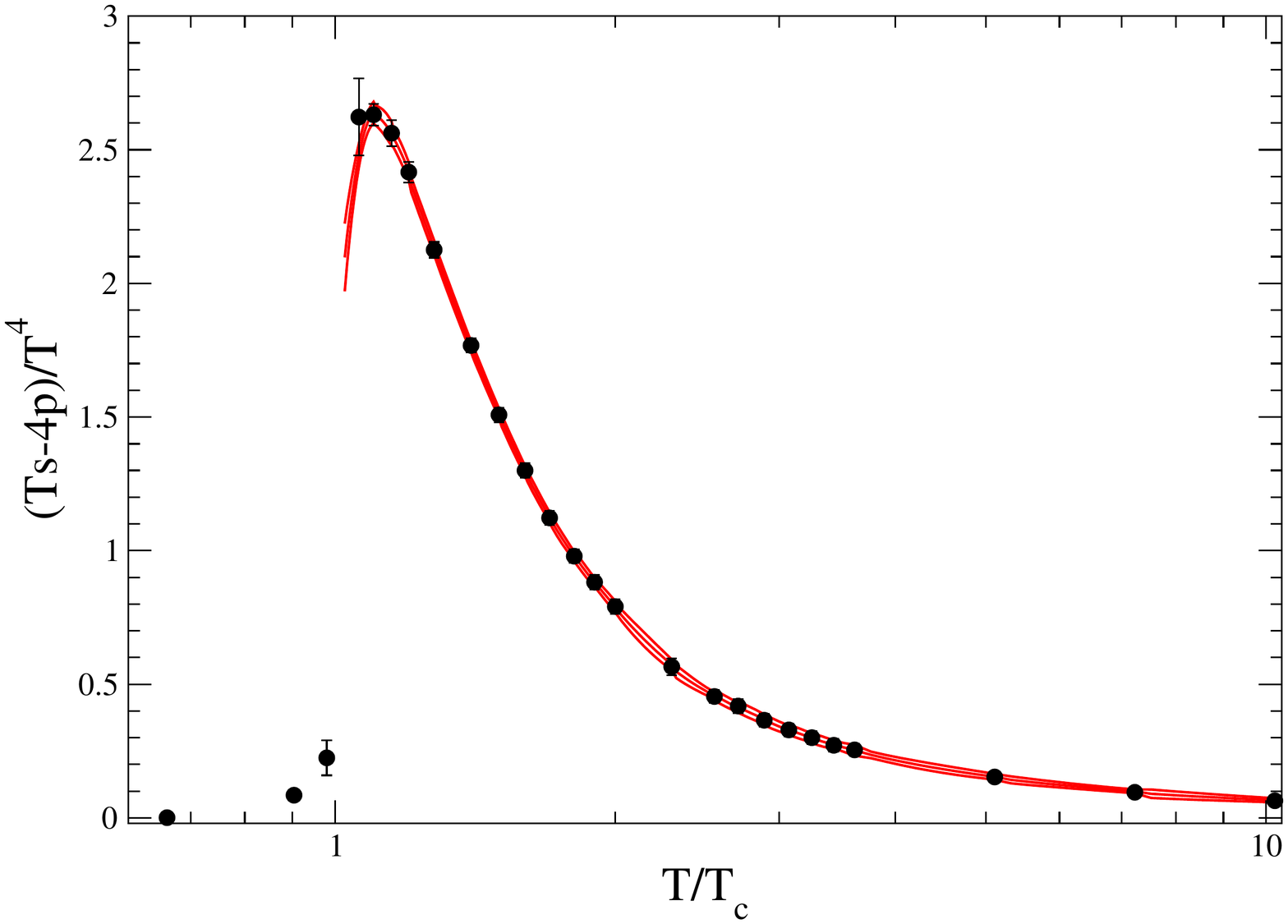}
\end{minipage}
\caption{The potentials $s/T^3$, $p/T^4$ and $\epsilon/T^4$ are shown as a
function of $T/T_c$ on the left, while on the right the anomaly
$(T s-4p)/T^4$ is plotted versus $T/T_c$.\label{fig:results}}
\end{center}
\end{figure*}
\indent In Figure~\ref{plotCL} we show the extrapolation to the continuum limit of
the tree-level improved definition of $s/T^3$ at
temperatures $T/T_c=1.2, 1.5$ and $2.0$. As expected, lattice artifacts turn out to be
very small. Results at other temperatures are qualitatively similar. Data generated
with $\vec\xi = (1,0,0)$ are also compatible with a constant behaviour, i.e. no
lattice spacing effects are observed within errors. Most of the uncertainty on $s/T^3$
comes from $Z_T$, since in our simulations $\langle T_{0k} \rangle_{\vec\xi}$ is measured
always with a statistical accuracy of permille or better.
We extrapolated to the continuum limit $s/T^3$ linearly in
$(a/L_0)^2$. When discretization effects are not visible, we also fit the data 
to a constant taking into account the correlation among the values of the renormalization
constant at different couplings. The fit are always similar to those
showed in Figure~\ref{plotCL}. All results are reported in Table~\ref{tab:continuum},
and shown in Figures~\ref{fig:sPT} and \ref{fig:results}. Mainly for the
computation of the pressure, see below, we have also carried out a
rough measurement of the entropy density at $T_c$ in the deconfined phase. 
The continuum extrapolation gives $s/T_c^3=1.70(24)$, where an estimate
of the systematics is included in the error. The analogous rough measurement in
the confined phase gives $s/T_c^3=0.37(15)$.\\
\indent Above a few $T_c$ and within the statistical errors, $s/T^3$ exhibits a striking linear
behaviour in $\ln(T/T_c)^{-1}$ over almost 2 orders of magnitude, see Figure~\ref{fig:sPT}.
For $T\to\infty$ data point straight to the Stefan-Boltzmann value, within errors, but with
a slope almost 5 times
smaller in magnitude than the leading-order perturbative prediction.

\section{The equation of state}
For $T\geq T_c$ the entropy density is well represented by the Pad\'e
interpolating formula 
\be\label{eq:pades}
\frac{s}{T^3} = 
\begin{cases}\displaystyle
  \frac{s_1 + s_2 w + s_3 w^2}{1 + s_4 w + s_5 w^2} & {\rm if}\;\;
  T/T_c \in [1.0,3.433]\, ,\\[0.375cm]
  s_6 +s_7 w^{-1}   & {\rm if}\;\;
   T/T_c \in [3.433,231.3]\, ,
\end{cases}
\ee
where $w=\ln(T/T_c)$, and the coefficients are:
$s_1= 1.7015$, $s_2= 77.757$, $s_3= 232.33$, $s_4= 19.033$, $s_5=32.200$, 
$s_6=6.9829$, $s_7=-1.0348$. The error attached to this curve at a given
point can be safely estimated by interpolating linearly those of the
two closest data points. This results in an uncertainty of $0.5\%$ for $T/T_c> 1.5$,
which increases up to $1\%$ going backward to $1.1$. For $T\rightarrow\infty$,
the formula reproduces the Stefan-Boltzmann value within errors.
Barring weird functional forms above $231.3$~$T_c$ which cannot be logically excluded,
Eq.~(\ref{eq:pades}) can be taken as a parametrization of the entropy for all temperatures
above $T_c$, with an error of $0.5\%$ for $T/T_c\geq 1.5$.\\
\indent Within the rather large errors that we have for $T\leq T_c$,
$\ln(s/T^3)$ can be fitted linearly in $T_c/T$ over the
four data points in this range. The quality of the fit is very good.\\[0.125cm] 
\indent Once the entropy density is known, the pressure is computed as
\be\label{eq:pT}
p(T) = \int_0^T\!\! s(T')\, dT'\; . 
\ee
For $T \leq T_c$, $p$ is computed by integrating the curve
resulting from the fit of $\ln(s/T^3)$ described above.
For $T\geq T_c$, the integral
is carried out as a sum
of integrals between each couple of consecutive points at which $s$
has been measured. Between any two such points,
$s/T^3$ is interpolated by a quadratic curve
in $\ln(T/T_c)$, with the three coefficients fixed by
fitting the four data points closest to the integration
region. The quality of the fits is always excellent, and the
distance of the curve from the data is always a small fraction
of the standard deviation. 
The statistical errors on $p$ are computed by propagating linearly
those on $s$, which in turn
are dominated by the ones on the renormalization constant. The
results for the pressure are reported in Table~\ref{tab:continuum} and shown in the
left plot of Figure~\ref{fig:results}. Being
the entropy a very smooth function, the
systematics due to the choice of the interpolating
function is negligible with respect to the statistical error.
As a further check, we have grouped the
data in three samples of consecutive points and fitted them
with independent Pad\'e interpolants. By integrating the three of
them in $T$, the results for the pressure
are in perfect agreement with those in Table~\ref{tab:continuum}.
Deviations are always a small fraction of the statistical
error.\\
\indent Analogously to the entropy density, for $T\geq T_c$ the values for the
pressure in Table~\ref{tab:continuum} are well
represented by the Pad\'e interpolating formula 
\be\label{eq:padep}
\frac{p}{T^4} = 
\begin{cases}\displaystyle
  \frac{p_1 + p_2 w + p_3 w^2}{1 + p_4 w + p_5 w^2} & {\rm if}\;\;
\displaystyle\frac{T}{T_c}\in[1.0,3.433]\\[0.375cm]
\displaystyle
\frac{p_6 +p_7 w^{-1}}{1 + p_8 w^{-1}}  & {\rm if}\;\;
\displaystyle\frac{T}{T_c}\in [3.433,231.3]
\end{cases}
\ee
where again $w=\ln(T/T_c)$, and the coefficients are:
$p_1= 0.022288$, $p_2= 2.0194$, $p_3= 10.030$, $p_4= 2.0941$, $p_5= 5.6006$, 
$p_6=1.7469$, $p_7=-0.79281$ and $p_8=-0.30484$. The error can be
computed as for the curve of the entropy. It corresponds
to $0.5\%$ for $T/T_c\geq 3$, and it increases 
up to $5\%$ going backward to $1.1$. For $T\rightarrow\infty$,
the formula reproduces the Stefan-Boltzmann value
within errors and can be taken as a parametrization of the pressure
for all temperatures above $T_c$, with an error of $0.5\%$ for $T/T_c\geq 3$.\\
\indent Once the entropy and the pressure are known, the energy density
is computed by using the relation
\be\label{eq:edens}
\frac{\epsilon}{T^4} = \frac{s}{T^3} - \frac{p}{T^4}\; . 
\ee
The results are again reported in Table~\ref{tab:continuum} and shown
in the left plot of Figure~\ref{fig:results}.  The central
values of $\epsilon$ can be computed by inserting
Eqs.~(\ref{eq:pades}) and (\ref{eq:padep}) into Eq.~(\ref{eq:edens}). The error is $0.5\%$
for $T/T_c\geq 1.3$ while it grows up to $0.7\%$ going backward to
$1.1$. The anomaly, computed as $(T s - 4p)/T^4$,
is shown in the right plot of Figure~\ref{fig:results}.\\
\indent It should be stressed once more that all results reported in this
section refer to the continuum theory.

\section{Discussion and conclusions}
At our largest temperature, approximatively $230\, T_c$, the entropy density still
differs from the Stefan-Boltzmann value by roughly $3\%$. It raises
linearly with $\ln(T/T_c)^{-1}$ following Eq.~(\ref{eq:pades}) which also connects well
the data to the Stefan-Boltzmann value at $T\rightarrow\infty$.\\
\indent The perturbative expression
is known to have a poor convergence rate~\cite{Kajantie:2002wa}. If we use
$\Lambda_{\msbar}\, r_0=0.586(48)$~\cite{Capitani:1998mq,Necco:2001xg},
this can be seen in Figure~\ref{fig:sPT} where the perturbative predictions at
the various orders are shown. It must be said that if one fixes the $O(g^6)$
undetermined coefficient so to be able to reproduce our results at large $T$,
a reasonable description of the data above $5 T_c$ or so is achieved. This comes
at the price of having an $O(g^6)$ contribution at $T= 231.3 T_c\sim 68$~GeV
which is roughly $50\%$ of the
total correction to the entropy density given by the other terms. The perturbative
formula is clearly of little help in determining the EoS of the theory in this interesting
temperature range. The (essentially) linear functional form followed by the data
in Figure~\ref{fig:sPT}, with a slope which is about 5 times smaller in magnitude
than the leading pertubative prediction, may require sophisticated theoretical tools
to be understood analytically, e.g.
Refs.~\cite{Braaten:1989mz,Blaizot:2003iq,Andersen:2004fp,Mogliacci:2013mca} and
references therein.\\
\indent For temperatures smaller than $T_c$ our results for the entropy density
agree with those in Refs.~\cite{Meyer:2009tq,Borsanyi:2012ve,Giusti:2014ila}. Between $T_c$ and
$3\, T_c$ they compare well with those in Refs.~\cite{Boyd:1996bx}, that have
significantly larger errors, and \cite{Giusti:2014ila} while we observe a significant disagreement with the
more precise ones in Ref.~\cite{Borsanyi:2012ve} as already reported in
Ref.~\cite{Giusti:2014ila}. In particular we find a discrepancy of  
$4$ to $6$ standard deviations between $T_c$ and $1.5\, T_c$, corresponding to
a $2$ to $4$ percent effect, which
becomes less than $2$ standard deviations above $3\, T_c$, see
Ref.~\cite{Pepe:2016noi} for a detailed discussion. For temperatures larger
than $3 T_c$, our results are in good agreement with Refs.~\cite{Boyd:1996bx,Borsanyi:2012ve,Giusti:2014ila}.
Similar discrepancies are propagated to pressure and energy density.\\
\indent To understand the origin of the disagreement, it is instructive to compare
directly the anomaly. This is the primary quantity computed in
Ref.~\cite{Borsanyi:2012ve} from which all other potentials are derived. 
Results for the anomaly in Refs.~\cite{Boyd:1996bx} and \cite{Borsanyi:2012ve} disagree
significantly, see Figure~2 and 4 in \cite{Borsanyi:2012ve} and ~\cite{Pepe:2016noi}
respectively.
Between $1.1 T_c$ and $1.4 T_c$, our results for $(T s-4p)/T^4$  confirm those in \cite{Boyd:1996bx}
while they disagree with \cite{Borsanyi:2012ve} by 2 to 4 standard deviations.  Our continuum
values, however, compare well with the finer lattice spacing results  in
Ref.~\cite{Borsanyi:2012ve}, i.e. before the continuum limit extrapolation is carried out.
We remark that similar disagreements with the data of Ref.~\cite{Borsanyi:2012ve} close to the
peak of the trace anomaly have been reported also in Refs.~\cite{Umeda:2014ula} and~\cite{Kitazawa:2016dsl}.
Above $1.4 T_c$ no significant discrepancy is observed for the anomaly.\\[0.125cm]
\indent The study reported here demonstrates that lattice gauge theory with
shifted boundary conditions offers a theoretically sound, simple and extremely powerful tool for an
accurate determination of the EoS over several orders of magnitude in the temperature.
Our results call for a non-perturbative determination of the EoS in QCD over an
analogous range of temperatures where perturbation theory cannot help.

\section{Acknowledgments}
Simulations have been performed on the BG/Q Fermi and on the
PC-clusters Galileo and Marconi at CINECA (CINECA-INFN and CINECA-Bicocca agreements),
and on the PC-cluster Wilson at Milano-Bicocca. We thankfully acknowledge the computer
resources and technical support provided by these institutions.

\appendix

\section{Numerical results}
In this appendix we collect the primary numerical results of this study.
For each lattice simulated, with bare gauge coupling $g_0^2$ and temporal
extension $L_0/a$, we report the expectation values of the space-time
component of the bare energy momentum tensor
$\langle T_{0k} \rangle_{\vec \xi} $ in Table~\ref{tab:matel}.

\begin{table*}[t!]
\begin{center}
\setlength{\tabcolsep}{.30pc}
\begin{tabular}{|r|cc|cc|cc|cc|cc|}
\hline%\\[-0.175cm]
& & & & & & & & & & \\[-0.25cm]  
$L_0/a$  & 
$6/g_0^2$ & $\langle T_{0k} \rangle_{\vec \xi} \times 10^{4} $&
$6/g_0^2$ & $\langle T_{0k} \rangle_{\vec \xi} \times 10^{4} $&
$6/g_0^2$ & $\langle T_{0k} \rangle_{\vec \xi} \times 10^{4} $&
$6/g_0^2$ & $\langle T_{0k} \rangle_{\vec \xi} \times 10^{4} $&
$6/g_0^2$ & $\langle T_{0k} \rangle_{\vec \xi} \times 10^{4} $\\[0.125cm]
\hline
\hline
& \multicolumn{2}{|c|}{$T= 0.980\, T_c$} 
& \multicolumn{2}{|c|}{$T= 1.061\, T_c$} 
& \multicolumn{2}{|c|}{$T= 1.100\, T_c$} 
& \multicolumn{2}{|c|}{$T= 1.150\, T_c$} 
& \multicolumn{2}{|c|}{$T= 1.200\, T_c$} \\
\hline
 5 &  6.1026  &     0.071(5) &  6.2515  &     0.502(3) &  6.0491  &     4.493(5)   &  6.0761  &     4.967(4)   &  6.1026 &      5.321(5)  \\
 6 &  6.2230  &     0.043(4) &  6.3845  &     0.251(5) &  6.1643  &     2.186(4)   &  6.1940  &     2.424(4)   &  6.2230 &      2.603(4)  \\
 7 &  6.3331  &     0.024(5) &  6.5031  &     0.141(5) &  6.2702  &     1.1993(28) &  6.3020  &     1.329(4)   &  6.3331 &      1.438(4)  \\
 8 &    -     &        -     &  6.6081  &     0.082(4) &  6.3678  &     0.7146(26) &  6.4012  &     0.798(4)   &  6.4337 &      0.859(4)  \\
10 &    -     &        -     &    -     &        -     &    -     &        -       &  6.5746  &     0.3372(18) &  6.6081 &      0.3626(18)\\
\hline
& \multicolumn{2}{|c|}{$T= 1.278\, T_c$} 
& \multicolumn{2}{|c|}{$T= 1.400\, T_c$} 
& \multicolumn{2}{|c|}{$T= 1.500\, T_c$} 
& \multicolumn{2}{|c|}{$T= 1.600\, T_c$} 
& \multicolumn{2}{|c|}{$T= 1.700\, T_c$} \\
\hline
 3  & 5.8506  &   44.508(22)  &   -     &      -      &   -     &      -      &   -     &      -      &   -     &      -       \\
 4  & 6.0056  &   13.889(7)   &   -     &      -      &   -     &      -      &   -     &      -      &   -     &      -       \\
 5  & 6.1429  &   5.745(10)   & 6.2037  &   6.255(8)  & 6.2515  &   6.554(9)  & 6.2975  &   6.789(4)  & 6.3418  &   7.010(7)   \\
 6  & 6.2670  &   2.812(4)    & 6.3331  &   3.058(3)  & 6.3845  &   3.218(5)  & 6.4337  &   3.341(4)  & 6.4805  &   3.443(4)   \\
 7  & 6.3798  &   1.5484(21)  & 6.4495  &   1.683(4)  & 6.5031  &   1.767(4)  & 6.5538  &   1.828(4)  & 6.6015  &   1.876(4)   \\
 8  & 6.4822  &   0.9209(22)  & 6.5538  &   1.005(4)  & 6.6081  &   1.048(3)  & 6.6588  &   1.089(4)  & 6.7062  &   1.116(4)   \\
10  & 6.6575  &   0.3904(16)  & 6.7286  &   0.4209(18)& 6.7815  &   0.4380(17)& 6.8300  &   0.4487(19)& 6.8745  &   0.4660(18) \\
\hline
& \multicolumn{2}{|c|}{$T= 1.807\, T_c$} 
& \multicolumn{2}{|c|}{$T= 1.900\, T_c$} 
& \multicolumn{2}{|c|}{$T= 2.000\, T_c$} 
& \multicolumn{2}{|c|}{$T= 2.300\, T_c$} 
& \multicolumn{2}{|c|}{$T= 2.556\, T_c$} \\
\hline
 3  & 6.0403  &  54.278(22)  &   -     &      -      &   -     &      -       &   -     &       -      &   -     &       -       \\
 4  & 6.2257  &  17.262(5)   &   -     &      -      &   -     &      -       &   -     &       -      &   -     &       -       \\
 5  & 6.3875  &   7.203(5)   & 6.4256  &   7.358(4)  & 6.4652  &   7.497(4)   & 6.8285  &    0.9919(22)& 6.6262  &    7.981(4)   \\
 6  & 6.5282  &   3.536(5)   & 6.5677  &   3.604(4)  & 6.6081  &   3.672(4)   & 6.9812  &    0.4859(23)& 6.7791  &    3.913(4)   \\
 7  & 6.6495  &   1.9369(21) & 6.6888  &   1.977(4)  & 6.7286  &   2.010(3)   & 7.1102  &    0.2684(10)& 6.9083  &    2.142(4)   \\
 8  & 6.7533  &   1.1458(18) & 6.7915  &   1.166(3)  & 6.8300  &   1.194(3)   & 7.2219  &    0.1576(8) & 7.0201  &    1.2694(27) \\
 9  &   -     &      -       &   -     &      -      & 6.9156  &   0.7502(26) &   -     &       -      &   -     &       -       \\
10  & 6.9183  &   0.4769(18) &   -     &      -      &   -     &      -       &   -     &       -      &   -     &       -       \\
\hline
& \multicolumn{2}{|c|}{$T= 2.711\, T_c$} 
& \multicolumn{2}{|c|}{$T= 2.891\, T_c$} 
& \multicolumn{2}{|c|}{$T= 3.072\, T_c$} 
& \multicolumn{2}{|c|}{$T= 3.253\, T_c$} 
& \multicolumn{2}{|c|}{$T= 3.433\, T_c$} \\
\hline
 5  & 6.6756  &   8.119(5) & 6.7297  &   8.251(6)  & 6.7806  &   8.352(5) & 6.8285  &   8.456(5) & 6.8738  &   8.536(4) \\
 6  & 6.8285  &   3.990(5) & 6.8826  &   4.032(5)  & 6.9334  &   4.085(5) & 6.9812  &   4.134(5) & 7.0265  &   4.181(4) \\
 7  & 6.9576  &   2.167(4) & 7.0117  &   2.204(4)  & 7.0624  &   2.236(4) & 7.1102  &   2.256(4) & 7.1555  &   2.271(3) \\
 8  & 7.0694  &   1.292(4) & 7.1234  &   1.304(4)  & 7.1741  &   1.318(4) & 7.2219  &   1.334(4) & 7.2671  &   1.349(4) \\
\hline
& \multicolumn{2}{|c|}{$T= 3.614\, T_c$} 
& \multicolumn{2}{|c|}{$T= 5.111\, T_c$} 
& \multicolumn{2}{|c|}{$T= 7.228\, T_c$} 
& \multicolumn{2}{|c|}{$T= 10.22\, T_c$} 
& \multicolumn{2}{|c|}{$T= 14.46\, T_c$} \\
\hline
 5  & 6.9168  &   8.627(4) & 7.2599  &   9.153(4) & 7.5499  &   9.501(4) & 7.8562  &   9.791(4) & 8.1424  &  10.030(6) \\
 6  & 7.0694  &   4.219(4) & 7.4120  &   4.456(4) & 7.7039  &   4.616(4) & 8.0060  &   4.747(5) & 8.2954  &   4.860(4) \\
 7  & 7.1984  &   2.299(4) & 7.5414  &   2.415(4) & 7.8349  &   2.509(4) & 8.1340  &   2.581(4) & 8.4260  &   2.640(4) \\
 8  & 7.3100  &   1.358(4) & 7.6541  &   1.428(4) & 7.9489  &   1.479(4) & 8.2458  &   1.522(4) & 8.5402  &   1.550(4) \\
\hline
& \multicolumn{2}{|c|}{$T= 20.44\, T_c$} 
& \multicolumn{2}{|c|}{$T= 28.91\, T_c$} 
& \multicolumn{2}{|c|}{$T= 40.89\, T_c$} 
& \multicolumn{2}{|c|}{$T= 57.82\, T_c$} 
& \multicolumn{2}{|c|}{$T= 81.78\, T_c$} \\
\hline
 5  & 8.4587  &  10.239(6)  & 8.7504  &  10.413(6) & 9.3518  &  10.710(6)    & 9.6387  &  10.844(5) & 9.9016  &  10.943(6) \\
 6  & 8.6123  &   4.961(14) & 8.9036  &   5.042(4) & 9.5022  &   5.178(4)    & 9.7913  &   5.238(4) & 10.0532 &   5.289(4) \\
 7  & 8.7419  &   2.688(4)  & 9.0330  &   2.729(4) & 9.6303  &   2.8053(27)  & 9.9212  &   2.842(4) & 10.1820 &   2.861(3) \\
 8  & 8.8542  &   1.578(3)  & 9.1449  &   1.611(4) & 9.7419  &   1.6507(24)  & 10.0344 &   1.668(4) & 10.2941 &   1.679(3) \\
\hline
& \multicolumn{2}{|c|}{$T= 115.6\, T_c$} 
& \multicolumn{2}{|c|}{$T= 163.6\, T_c$} 
& \multicolumn{2}{|c|}{$T= 231.3\, T_c$} 
& \multicolumn{2}{c}{} 
& \multicolumn{2}{c}{} \\
%\hline
\cline{1-7}
 5  & 10.1905 &  11.042(4)  & 10.6048 &  11.177(4) & 10.8883 &  11.270(4) & \multicolumn{4}{c}{} \\
 6  & 10.3436 &   5.340(4)  & 10.7531 &   5.402(4) & 11.0400 &   5.436(4) & \multicolumn{4}{c}{} \\
 7  & 10.4736 &   2.889(4)  & 10.8799 &   2.921(4) & 11.1697 &   2.942(4) & \multicolumn{4}{c}{} \\
 8  & 10.5867 &   1.699(4)  & 10.9908 &   1.717(4) & 11.2831 &   1.733(4) & \multicolumn{4}{c}{} \\
%\hline
\cline{1-7}
\end{tabular}
\caption{
  \label{tab:matel} Expectation values of the space-time component of the energy momentum
  tensor, $\langle T_{0k} \rangle_{\vec \xi} $, at finite lattice spacing. For each
  physical temperature, expressed in units of $T_c$, we report the value of the bare gauge
  coupling, $6/g_0^2$, and the lattice size in the temporal direction, $L_0/a$, considered in the Monte Carlo simulations.}
\end{center}
\end{table*}

\bibliographystyle{elsarticle-num}
\bibliography{EoS_revised}
\end{document}